\documentclass[showpacs,prd,twocolumn,floatfix,
nofootinbib,superscriptaddress]{revtex4}
\usepackage{graphicx}

\newcommand{\be}{\begin{equation}}
\newcommand{\ee}{\end{equation}}

\newcommand{\psibar}{\overline{\Psi}}

\begin{document}


\title{The $B$ and $B_s$ Meson Decay Constants from Lattice QCD}

\author{Heechang Na} \thanks{current address: Argonne Leadership Computing Facility,\\
Argonne National Laboratory, 
Argonne, IL 60439, USA.}
\affiliation{Department of Physics,
The Ohio State University, Columbus, OH 43210, USA}
\author{Chris J.\ Monahan}
\affiliation{Department of Physics, College of William and Mary, VA 23187-8795, USA}
\author{Christine T.\ H.\ Davies}
\affiliation{SUPA, School of Physics \& Astronomy,
University of Glasgow, Glasgow, G12 8QQ, UK}
\author{Ron Horgan}
\affiliation{DAMTP, Cambridge University, CB3 0WA, UK}
\author{{G.Peter} Lepage}
\affiliation{Laboratory of Elementary Particle Physics,
Cornell University, Ithaca, NY 14853, USA}
\author{Junko Shigemitsu}
\affiliation{Department of Physics,
The Ohio State University, Columbus, OH 43210, USA}

\collaboration{HPQCD Collaboration}
\noaffiliation


\begin{abstract}
We present a new determination of the $B$ and $B_s$ meson decay constants 
using NRQCD $b$-quarks, HISQ light and strange valence quarks and 
the MILC collaboration $N_f = 2 + 1$ lattices. The new calculations improve on 
HPQCD's earlier work with NRQCD $b$-quarks  
by replacing AsqTad with HISQ valence quarks, by including 
a more chiral MILC fine ensemble in the analysis, 
and by employing better tuned quark masses and 
 overall scale.
We find $f_B = 0.191(9)$GeV, $f_{B_s} = 0.228(10)$GeV and $f_{B_s}/f_B =1.188(18)$.  
Combining the new value for 
$f_{B_s}/f_B$ with a recent very precise determination of the $B_s$ meson  
decay constant based on HISQ $b$-quarks, 
$f_{B_s} = 0.225(4)$GeV,  leads to
$f_B = 0.189(4)$GeV. With errors of just 2.1\% this represents the most precise
 $f_B$ available today.

\end{abstract}


\maketitle


\section{Introduction}
Precision electroweak data gathered at the $B$ factories, the Tevatron and at 
LHCb is allowing particle physicists to carry out stringent tests of the Standard Model 
(SM) and search for hints of New Physics (NP). Several groups, for instance, are 
studying global fits to the Cabibbo-Kobayashi-Maskawa (CKM) unitarity triangle (UT)
and checking whether various combinations of constraints coming from 
experiment and theory can be accommodated consistently with each other 
\cite{ls1,ls2,llw}. 
In recent years some 
tensions at the 2-3 $\sigma$ level within the SM have emerged from these studies and 
it will be very interesting to see whether future improvements in experimental 
and theory inputs will remove these tensions or conversely elevate them to 
serious hints of NP.

Lattice QCD is playing an important role in UT analyses, providing 
crucial inputs such as $\epsilon_K$, $\hat{B}_{B_q}$, $\xi = f_{B_s} \sqrt{B_{B_s}}/
f_B \sqrt{B_B}$, $f_B$ and information on semileptonic form factors \cite{reviews}.
  To make progress 
in resolving the tensions in UT analyses it is imperative to reduce 
 the errors in current lattice results. 
In reference \cite{ls1} the $B$ meson decay constant $f_B$ is not used as an 
input for the global fits but becomes instead one of the fit outputs 
$f_B^{(fit)}$.  This $f_B^{(fit)}$ is then compared with the SM (i.e. Lattice QCD) 
value $f_B^{(QCD)}$  to check for consistency. 
The authors of reference \cite{ls1} experiment with dropping different processes in 
their global fits and study how this affects $f_B^{(fit)}$ 
and when $f_B^{(fit)}$ agrees best with $f_B^{(QCD)}$. 
Using this fit-comparison procedure, the authors attempt
 to determine the  dominant source of deviations from 
the SM, e.g. whether it is coming from
$B \rightarrow \psi K_s$ (sin$2\beta$),
 $B_s$ and $B_d$ mixing, Kaon mixing ($\epsilon_K$) or $B \rightarrow \tau \nu$.
Needless to say this $f_B^{(fit)} - f_B^{(QCD)}$ comparison method requires knowing 
 $f_B^{(QCD)}$ as accurately as possible. In this article 
we significantly reduce errors in $f_B^{(QCD)}$.
With reduced errors, the $B$ meson decay constant will hopefully help 
further constrain UT analyses in the future. 

In the next section we introduce the lattice setup  
 and explain how the bottom and strange 
quark masses were fixed in our lattice actions.
  Section III discusses 
operator matching between heavy-light currents in full continuum QCD and in 
the lattice theory. We describe two-point correlators and the smearings 
employed. In section IV we present our fitting strategies to the 
two-point correlators and describe how the extracted amplitudes lead 
to the hadronic matrix elements relevant for determining decay constants. 
This section also includes summary tables of fit results for 
$\Phi_s = f_{B_s} \sqrt{M_{B_s}}$, $\Phi = f_B \sqrt{M_B}$, 
and their ratios for each of the 6 MILC ensembles that we work with.  
Then in section V we explain how continuum and chiral limit physics 
is extracted from our simulation data. 
 Section VI discusses 
results at the physical point and the error budget and we conclude 
with a summary in section VII.  
For the rest of this article we omit the ``QCD'' in $f_B^{(QCD)}$.
\section{ The Lattice Setup and Tuning of Bare Quark Masses}

HPQCD's previous work on $B$ and $B_s$ meson decay constants with  
NRQCD $b$-quarks used AsqTad light and strange quarks \cite{bmixing}. 
 It utilized the 
MILC AsqTad $N_f = 2 + 1$ lattices \cite{milc}. In the present work we replace 
the AsqTad valence quarks by their Highly Improved Staggered Quark (HISQ) \cite{hisq}
counterparts thereby reducing the dominant discretization errors 
coming from staggered taste breaking by roughly a factor of three \cite{mixed}. 
Details of 
the MILC ensembles employed here are given in Table I.   There is considerable 
overlap between the MILC ensembles used in the present article and in 
\cite{bmixing}.  In \cite{bmixing}
 an additional coarse ensemble with sea quark masses $m_l/m_s = 
0.007/0.05$ was employed.  Here we have added instead a third, more-chiral fine 
ensemble, the $40^3 \times 96$ Set F0 with $m_l/m_s = 0.0031/0.031$.

For the $b$-quarks in our simulations we use the same NRQCD action employed 
in \cite{bmixing}.  
Since the publication of \cite{bmixing} the HPQCD collaboration has updated 
the value of the scale parameter $r_1$ to $r_1 = 0.3133(23)$fm \cite{r1}, 
and this necessitated 
a retuning of all quark masses including the bare $b$-quark mass $aM_b$ for 
all MILC ensembles in Table I. 
 To fix  $aM_b$ we use the spin averaged $\Upsilon$ mass. One calculates,
\be
\overline{M}_{b \overline{b}}
 \equiv \frac{1}{4} \left [ 3 M_{kin}(^3S_1) + M_{kin}(^1S_0) \right ],
\ee
with 
\be
M_{kin} = \frac{p^2 - \Delta E_p^2}{2 \Delta E_p},
\qquad \qquad \Delta E_p = E(p) - E(0),
\ee
and compares with the experimental value
 (adjusted 
for the absence of electromagnetic, annihilation and sea charm quark 
effects in our simulations) of 9.450(4)GeV \cite{gregory}.  
Results from this tuning are shown in Fig.1.   Errors in the data points 
include statistical and $r_1/a$ errors.  One sees that these are much 
smaller than the 0.7\% error in the absolute physical value of $r_1$.  
To achieve such small statistical errors in $M_{kin}$ it was crucial to 
employ random wall sources for the NRQCD $b$-quark propagators.
Most of the tuning of $aM_b$ was carried out with momentum $2 \pi /(aL)$ for 
ensembles C1, C2, C3, F1 and F2, and with momentum $4 \pi/(aL)$ for 
ensemble F0. However, we have checked on one ensemble that 
consistent $M_{kin}$ values result from higher (but not too large) momenta 
as well.   For instance on C2 with $aM_b = 2.8$ (slightly larger than the 
actual physical $b$-quark mass) one finds $aM_{kin}(^3S_1) = 5.933(15)$ for 
momentum $2 \pi/(aL)$ and 
$aM_{kin}(^3S_1) = 5.941(15)$ for momentum $4 \pi/(aL)$.

\begin{table}
\caption{
Simulation details on three ``coarse'' and three ``fine'' MILC ensembles.
}
\begin{tabular}{|c|c|c|c|c|c|}
\hline
Set &  $r_1/a$ & $m_l/m_s$ (sea)   &  $N_{conf}$&
 $N_{tsrc}$ & $L^3 \times N_t$ \\
\hline
\hline
C1  & 2.647 & 0.005/0.050   & 1200  &  2 & $24^3 \times 64$ \\
\hline
C2  & 2.618 & 0.010/0.050  & 1200   & 2 & $20^3 \times 64$ \\
\hline
C3  & 2.644 & 0.020/0.050  &  600  & 2 & $20^3 \times 64$ \\
\hline
\hline
 F0  & 3.695  &  0.0031/0.031  & 600  & 4 & $40^3 \times 96$ \\
\hline
F1  & 3.699 & 0.0062/0.031  & 1200  & 4  & $28^3 \times 96$ \\
\hline
F2  & 3.712 & 0.0124/0.031  & 600  & 4 & $28^3 \times 96$ \\
\hline
\end{tabular}
\end{table}

The $s$-quark mass was tuned to the (fictitious) $\eta_s$ mass of 
0.6858(40)GeV \cite{r1}. Fig.2 shows results for this tuning. 
All but the Set F0 point (most chiral point on plot) were fixed already in \cite{dtok}. 
Having fixed the bottom and strange quark masses on each 
ensemble one can investigate the mass combination 
$M_{B_s} - \overline{M}_{b \overline{b}}/2$.
 The leading dependence on the heavy quark 
mass cancels in this difference, so one is testing how well the lattice 
actions are simulating QCD boundstate dynamics.  Results for this mass 
difference are shown in Fig.3.  Within the $r_1$ scale error and additional 
$\sim$10MeV uncertainty from relativistic corrections to 
$\overline{M}_{b \overline{b}}$ one sees agreement with experiment
 after removing discretization effects.

Table II summarizes the valence quark masses used in this article.  
We include the HISQ valence charm quark masses for each ensemble, 
since these provide a convenient scale in the chiral 
extrapolations of section V.  The charm quark masses were fixed by tuning 
to the $\eta_c$ mass.
The light HISQ valence quark mass $m_l$ is chosen so that 
$m_l(valence)/m_s(valence)$ is close to $m_l(sea)/m^{phys}_{s,AsqTad}$, 
where $m^{phys}_{s,AsqTad}$ corresponds to the physical AsqTad strange quark 
mass.
As a final consistency check of our lattice setup, 
we have looked at the $B_s-B$ mass difference.  This is shown in Fig.4 .

\begin{figure}
\includegraphics*[width=8.0cm,height=9.0cm,angle=270]{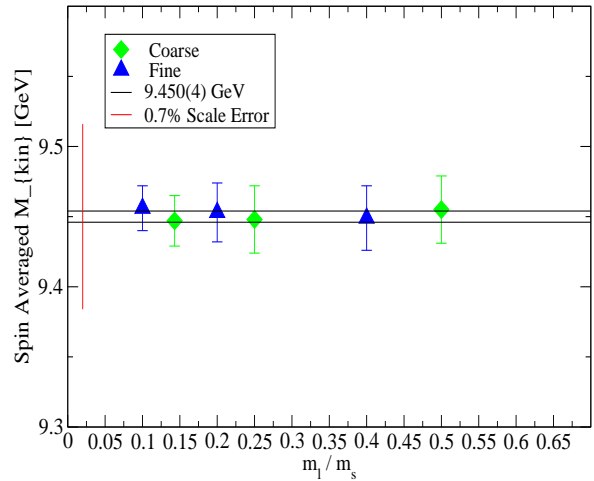}
\caption{
Tuning of the $b$-quark mass via the spin averaged $\Upsilon$ mass. 
9.450GeV corresponds to the experimental value adjusted for 
lack of electromagnetic, annihilation and sea charm quark effects in the 
simulations.
 }
\end{figure}
\begin{figure}
\includegraphics*[width=8.0cm,height=9.0cm,angle=270]{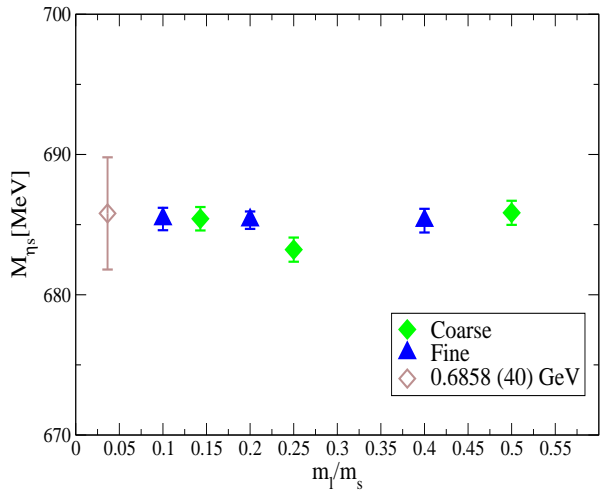}
\caption{
Tuning of the strange quark mass via the fictitious $\eta_s$ meson.\\
 }
\end{figure}

\begin{table}
\caption{
Valence quark masses
}
\begin{tabular}{|c|c|c|c|c|}
\hline
Set & $a m_l$  & $a m_s$ & $a m_c$ &  $aM_b$  \\
\hline
\hline
C1  &  0.0070 &  0.0489 & 0.6207  & 2.650  \\
\hline
C2  & 0.0123 & 0.0492  &  0.6300 & 2.688  \\
\hline
C3  & 0.0246 & 0.0491 & 0.6235 & 2.650  \\
\hline
\hline
 F0  &  0.00339  & 0.0339 & 0.4130 & 1.832  \\
\hline
F1  & 0.00674 & 0.0337 & 0.4130 & 1.832  \\
\hline
F2  & 0.0135  & 0.0336 & 0.4120 & 1.826  \\
\hline
\end{tabular}
\end{table}

\begin{figure}
\includegraphics*[width=8.0cm,height=9.0cm,angle=270]{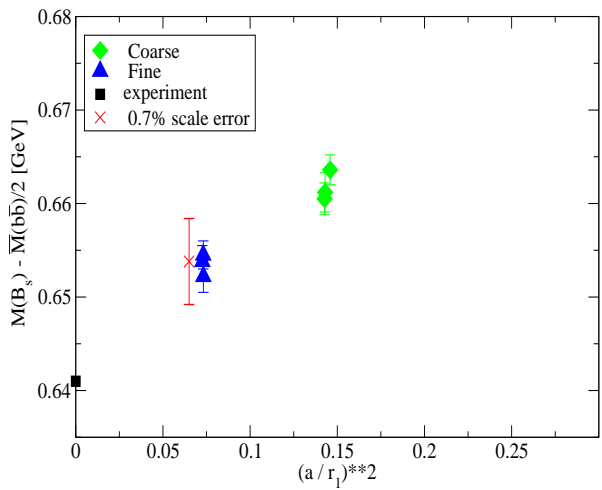}
\caption{
The mass difference $M_{B_s} - \overline{M}_{b \overline{b}} / 2$.
 }
\end{figure}
\begin{figure}
\includegraphics*[width=8.0cm,height=9.0cm,angle=270]{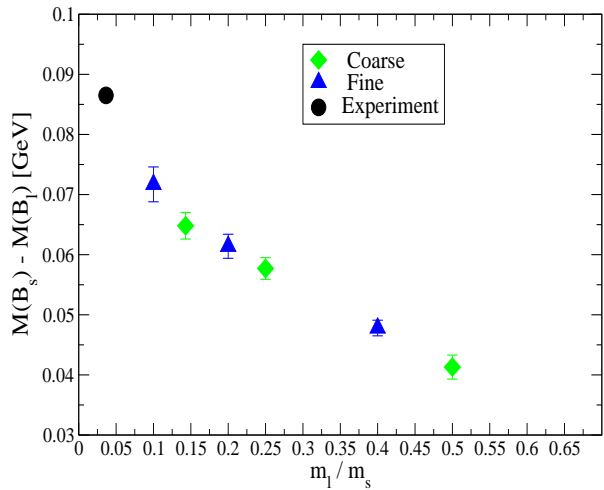}
\caption{
The $B_s$ - $B$ mass difference $\Delta M$ versus the light valence 
quark mass.
 }
\end{figure}

\section{Operator Matching and Relevant Correlators}
Decay constants $f_{B_q}$ are determined by calculating the 
matrix element of the heavy-light axial vector current $A_\mu$ 
 between the $B_q$ meson and  hadronic vacuum states.  For the temporal
component in the $B_q$ restframe one has,
\be
 \langle 0 | \; A_0 \; | B_q \rangle_{QCD} = M_{B_q} \; f_{B_q} .
\ee
Simulations are carried out with effective lattice theory currents,
\begin{eqnarray}
 J^{(0)}_{0}(x) & = & \psibar_q \,\Gamma_0\, \Psi_Q, \\
 J^{(1)}_{0}(x) & = & \frac{-1}{2M_b} \psibar_q
    \,\Gamma_0\,\mbox{\boldmath$\gamma\!\cdot\!\nabla$} \, \Psi_Q, \\
 J^{(2)}_{0}(x) & = & \frac{-1}{2M_b}  \psibar_q
    \,\mbox{\boldmath$\gamma\!\cdot\!\overleftarrow{\nabla}$}
    \,\gamma_0\ \Gamma_0\, \Psi_Q,
\end{eqnarray}
with $\Gamma_0 = \gamma_5 \gamma_0$ for decay constant calculations.
$\Psi_q$ is the HISQ action light or strange quark field 
(in its four component ``naive fermion'' form) and 
$\Psi_Q$ the heavy quark field with the upper two components given by the 
two-component NRQCD fields and the lower two components set equal to zero.
We have matched these effective theory currents to $A_0$ in full QCD at 
one-loop through order 
 $\alpha_s, \; \frac{\Lambda_{QCD}}{M}, \;
\frac{\alpha_s}{aM},\; a \alpha_s, \; \alpha_s \frac{\Lambda_{QCD}}{M}$. 
Details of the matching of NRQCD/HISQ currents will be presented in a 
separate publication \cite{perturb}.
 The calculations follow the strategy developed in 
\cite{cmjs} and employed for NRQCD/AsqTad currents in \cite{gulez}. One finds,
\begin{eqnarray}
\label{a0}
\langle A_0 \rangle_{QCD}  &=& ( 1 + \alpha_s
 \, 
\rho_0)\,\langle J^{(0)}_0 \rangle  + 
 (1 + \alpha_s   \,  \rho_1) \, \langle
J^{(1),sub}_0 \rangle \nonumber \\
 & & + \alpha_s  \,
 \rho_2 \, \langle J^{(2),sub}_0 \rangle,
   \\
 &&  \nonumber \\
\label{jsub}
 J_0^{(i),sub} &=& J_0^{(i)} - \alpha_s \,   \zeta_{10}
J_0^{(0)} .
\end{eqnarray}
Here $\rho_0$, $\rho_1$, $\rho_2$ and $\zeta_{10}$ are the one-loop 
matching coefficients.

We use smeared heavy-light bilinears to represent the $B_q$ mesons. 
For instance, we create a meson at time $t_0$ via,
\be
\Phi_\alpha(\vec{x},t_0) \equiv \sum_{\vec{x}_1} \psibar_Q(\vec{x}_1,t_0) \sigma_\alpha
(\vec{x}_1 - \vec{x}) \Gamma_{sc} \Psi_q(\vec{x},t_0),
\ee
with $\Gamma_{sc} = \gamma_5$. For the smearing functions $\sigma_\alpha(\vec{x}_1 - 
\vec{x})$
 we use a $\delta$-function local smearing ($\alpha = 1$) 
or Gaussian smearings $ \propto e^{-|\vec{x}_1 - \vec{x}|^2/
(2 r_0^2)}$for two different widths $r_0$ and normalized to one ($\alpha = 2,3$).
We then calculate a $3 \times 3$ matrix of zero momentum  meson correlators with all 
combinations of source and sink smearings,
\be
\label{bcorrel}
C^B_{\beta,\alpha}(t,t_0) = \frac{1}{V} \sum_{\vec{x}} \sum_{\vec{y}} 
\langle \Phi^\dagger_\beta(\vec{y},t) \, \Phi_\alpha(\vec{x},t_0) \rangle,
\ee
with $V = L^3$. We use Gaussian widths in lattice units of size 
$r_0 = 3$ or $5$ on coarse ensembles and $r_0 = 4$ or $7$ on the fine ensembles.  
In addition to this matrix of $B$ correlators we also need correlators with 
$\Phi_\alpha$ at the source and $J_0^{(i)}$ at the sink for $i=0,1,2$,
\be
\label{jcorrel}
C^{Ji}_\alpha(t,t_0) = \frac{1}{V} \sum_{\vec{x}} \sum_{\vec{y}} 
\langle J_0^{(i)}(\vec{y},t) \, \Phi_\alpha(\vec{x},t_0) \rangle.
\ee
Since $\gamma_0 \Psi_Q = \Psi_Q$ it turns out that,
\be
C^{J0}_\alpha \equiv C^B_{\beta=1,\alpha}.
\ee
Furthermore for zero momentum correlators one can show that,
\be
C^{J2}_\alpha \equiv C^{J1}_\alpha ,
\ee
so only the three $C^{J1}_\alpha$, $\alpha = 1,2,3$, 
 are required in addition to the $3 \times 3$ matrix $C^B_{\beta,\alpha}$.

The spatial sums $\sum_{\vec{y}}$ in (\ref{bcorrel}) and (\ref{jcorrel}) are done 
at the sink, and so can be handled very easily.  We implement the $\sum_{\vec{x}}$ 
sums at the source via random wall sources. This is described for instance  in 
reference \cite{gregory}. Here we give some of the  explicit formulas. 
In terms of quark propagators for the $\Psi_Q$ and $\Psi_q$ fields eq.(\ref{bcorrel}) 
becomes (we set $t_0 = 0$ for simplicity),
\begin{eqnarray}
& & C^B_{\beta,\alpha}(t) = 
 \frac{1}{V} \sum_{\vec{x}} \sum_{\vec{y}}
\sum_{\vec{x}_1} \sum_{\vec{y}_1}  \nonumber \\
 & & \langle tr \left \{ G_Q(\vec{y}_1 - \vec{x}_1,t)
 \sigma_\alpha(\vec{x}_1 - \vec{x}) 
\Gamma_{sc} G_q(\vec{x} - \vec{y},-t) \right. \nonumber \\
& & \qquad \qquad \qquad \qquad \qquad \left. 
  \Gamma_{sk} \sigma_\beta(\vec{y}_1 - 
\vec{y}) \right \} \rangle \nonumber \\
 &&  = \frac{1}{V} \sum_{\vec{x}} \sum_{\vec{y}}
\sum_{\vec{x}_1} \sum_{\vec{y}_1} \langle tr \left \{ G_Q(\vec{y}_1 - 
\vec{x}_1,t) \right.
 \nonumber \\
  & &\left. 
\sigma_\alpha(\vec{x}_1 - \vec{x}) 
\Gamma_{sc} \gamma_5 G^\dagger_q(\vec{y} - \vec{x}, t) \gamma_5 
 \Gamma_{sk} \sigma_\beta(\vec{y}_1 - 
\vec{y}) \right \} \rangle. \nonumber \\
\end{eqnarray}
We set,
\be
\label{hgsm}
 G_Q^{(sm\alpha)}(\vec{y}_1, \vec{x},t) \equiv 
\sum_{\vec{x}_1} G_Q(\vec{y}_1 - \vec{x}_1,t) \sigma_\alpha(\vec{x}_1 - \vec{x}), 
\ee
and recall the relation between the naive HISQ propagator $G_q(y-x)$ 
and the one component HISQ quark propagator $G_\chi(y-x)$ \cite{wingate},
\be
 G_q(y - x) = \Omega(y) G_\chi(y-x) \Omega^\dagger(x),
\ee
or equivalently,
\be
 G^\dagger_q(y-x ) = \Omega(x)
 [\Omega(y) G_\chi(y-x)]^\dagger,
\ee
with,
\be 
\Omega(x) \equiv \gamma_0^{x_0} \gamma_1^{x_1}\gamma_2^{x_2}\gamma_3^{x_3}.
\ee
Setting $\Gamma_{sc} = \Gamma_{sk} = \gamma_5$ one has,
\begin{eqnarray}
\label{cb1}
& & C^B_{\beta,\alpha}(t) =  \nonumber \\
& &  \frac{1}{V} \sum_{\vec{x}} \sum_{\vec{y}}
 \sum_{\vec{y}_1} 
 \langle tr \left \{[ G^{(sm\alpha)}_Q(\vec{y}_1,\vec{x},t)
\Omega(x)] \right. \nonumber \\
& &  \qquad \qquad \left. [\Omega(y)G_\chi(\vec{y} - \vec{x}, t)]^\dagger
 \sigma_\beta(\vec{y}_1 - 
\vec{y}) \right \} \rangle  .
\end{eqnarray}
We introduce a random U(1) field $\xi(\vec{x})$ at each spatial site of the 
source time slice (in practice we employ separate U(1) fields for each 
color but suppress this index in the formulas given below) and replace,
\be
\frac{1}{V} \sum_{\vec{x}} \longrightarrow  \frac{1}{V} 
\sum_{\vec{x}} \sum_{\vec{x}^\prime} \xi(\vec{x}) \xi^\dagger(\vec{x}^\prime).
\ee
Eq.(\ref{cb1}) becomes,
\begin{eqnarray}
\label{cb2}
& & C^B_{\beta,\alpha}(t) =  \nonumber \\
& &  \frac{1}{V} \sum_{\vec{x}} \sum_{\vec{x}^\prime} \sum_{\vec{y}}
 \sum_{\vec{y}_1} 
 \langle tr \left \{[ G^{(sm\alpha)}_Q(\vec{y}_1,\vec{x},t)
\Omega(x) \xi(\vec{x})] \right. \nonumber \\
& &  \qquad \qquad \left. [\Omega(y)G_\chi(\vec{y} - \vec{x}^\prime, t)
 \xi(\vec{x}^\prime)]^\dagger
 \sigma_\beta(\vec{y}_1 - 
\vec{y}) \right \} \rangle.  \nonumber \\
\end{eqnarray}
An even more concise expression can be obtained if one defines,
\begin{eqnarray}
\label{hcorrel}
& &G^{(sm\alpha,rw)}_Q(\vec{y}_1,t) \equiv \frac{1}{\sqrt{V}} \sum_{\vec{x}} 
G^{(sm\alpha)}_Q(\vec{y}_1,\vec{x},t) \Omega(x)\xi(\vec{x}) \nonumber \\
& & =  \frac{1}{\sqrt{V}}\sum_{\vec{x}_1} 
 G_Q(\vec{y}_1 - \vec{x}_1,t) \sum_{\vec{x}}
\sigma_\alpha(\vec{x}_1 - \vec{x}) \Omega(x) \xi(\vec{x}), \nonumber \\
\end{eqnarray}
and,
\begin{eqnarray}
\label{lcorrel}
G^{(rw)}_q(\vec{y},t) & \equiv & \Omega(y) G^{(rw)}_\chi(\vec{y},t) \nonumber \\
& \equiv  &
\frac{1}{\sqrt{V}} \sum_{\vec{x}^\prime} 
\Omega(y) G_\chi(\vec{y} - \vec{x}^\prime,t)\xi(\vec{x}^\prime) . \nonumber \\
\end{eqnarray}
This leads to,
\begin{eqnarray}
\label{cb3}
& & C^B_{\beta,\alpha}(t) =  
 \sum_{\vec{y}} \sum_{\vec{y}_1}  \nonumber \\
 & & \langle tr \left \{[ G^{(sm\alpha,rw)}_Q(\vec{y}_1,t)]
 [G_q^{(rw)}(\vec{y}, t)]^\dagger
 \sigma_\beta(\vec{y}_1 - 
\vec{y}) \right \} \rangle.  \nonumber \\
\end{eqnarray}
Equations  (\ref{hcorrel}) and (\ref{lcorrel}) tell us 
that we should create NRQCD propagators with source,
\be
SC^\alpha_Q(\vec{x}_1) = \frac{1}{\sqrt{V}} \sum_{\vec{x}} \sigma_\alpha(\vec{x}_1 
-\vec{x}) \Omega(x) \xi(\vec{x}),
\ee
and HISQ propagators with source,
\be
SC_q(\vec{x}^\prime) = \frac{1}{\sqrt{V}} \xi(\vec{x}^\prime).
\ee
The double sum in (\ref{cb3}) is carried out via Fast Fourier Transforms.

\begin{figure}
\begin{center}
\includegraphics*[width=8.0cm,height=12.0cm,angle=270]{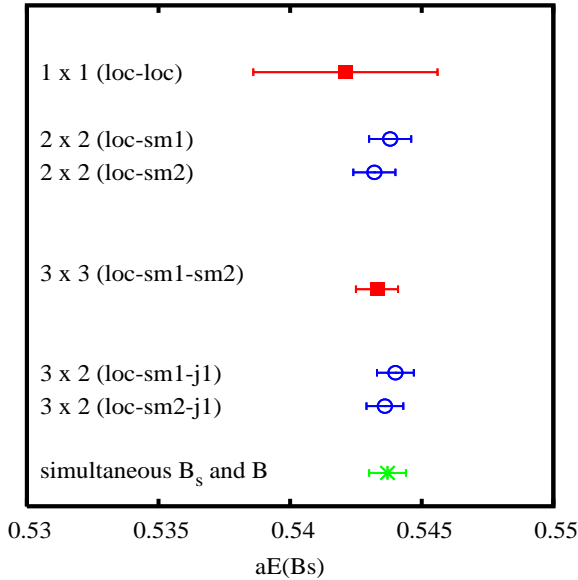}
\end{center}
\caption{
Examples of results from different matrix fits for the $B_s$ 
meson energy in lattice units.  These fit results are taken from 
the C2 ensemble and used $N=\tilde{N} = 7$ in eq.(\ref{twopnt}).
 }
\end{figure}

\section{ Fits and Data Analysis}
The $3 \times 3$ matrix of correlators $C^B_{\beta,\alpha}$ of eq.(\ref{bcorrel}) and 
the $C^{Ji}_\alpha$ of eq.(\ref{jcorrel}) for $ i = 1$ can be combined into a 
$4 \times 3$ matrix of correlators $C_{\beta,\alpha}$ with $C_{\beta,\alpha} \equiv 
C^B_{\beta,\alpha}$ for $\alpha, \beta = 1,2,3$ and 
$C_{\beta=4,\alpha=1,2,3} 
\equiv C^{J1}_{\alpha = 1,2,3}$. Various subsets of these correlators are then 
fit simultaneously to the form,
\begin{eqnarray}
\label{twopnt}
 && C_{\beta, \alpha}(t) \nonumber \\
 && =\sum_{j=0}^{N-1} b^\beta_j b^\alpha_j e^{-E_j (t-1)}
 + (-1)^t \sum_{k=0}^{\tilde{N}-1} 
\tilde{b}^\beta_k \tilde{b}^\alpha_k e^{-\tilde{E}_k (t-1)}, \nonumber \\
\end{eqnarray}
to extract the ground state energy $E_0$ and amplitudes $b^\beta_0$.
The hadronic matrix elements appearing in (\ref{a0}) are related to the 
amplitudes $b_0^\beta$ as,
\be
a^2 \, \langle 0 | J_0^{(0)}| B_q \rangle = \sqrt{2 M_{B_q} a} \; b^{\beta=1}_0, 
\ee
and
\be
a^2 \, \langle 0 | J_0^{(1)}| B_q \rangle = 
a^2 \, \langle 0 | J_0^{(2)}| B_q \rangle = 
\sqrt{2 M_{B_q} a} \; b^{\beta=4}_0.
\ee
The factors of $\sqrt{2 M_{B_q} a}$ come about due to differences in normalization 
of states in the effective lattice theory compared to the standard relativistic 
normalization of states.

We have investigated fits to various subsets of correlators (submatrices) 
taken from the full $4 \times 3$ matrix of 12 correlators.  For each correlator
 we fit data between $t = t_{min}$ and $t = t_{max}$ with 
$t_{min} = 2 \sim 4$ and $t_{max} = 16$ on coarse ensembles and 
$t_{min} = 4 \sim 8$ and $t_{max} = 24$ on the fine ensembles. 
In Fig.5 we show results for 
the $B_s$ energy in lattice units, $a E_{B_s}$, from fits to ensemble C2.  One sees 
 a large improvement upon going from a fit to a single local-local 
($\alpha, \beta = 1$) correlator to a $ 2 \times 2$ matrix of correlators 
($\alpha, \beta = 1,2$ or $\alpha, \beta = 1,3$).  There appears to be little 
further improvement when one goes to $3 \times 3$ matrices.  Our final fit 
results are taken from $3 \times 2$ matrix fits with $\alpha = 1,3$ and $\beta = 1,3,4$.
We do simultaneously a $ 3 \times 2$ fit  to $B$ correlators together 
with a $3 \times 2$ fit to $B_s$ correlators.  This allows us to get ratios such as 
$f_{B_s} \sqrt{M_{B_s}} /f_B \sqrt{M_B}$ and mass differences such as $M_{B_s} - M_B$ 
in a single fit with correctly correlated errors, in addition to the separate 
quantities $f_B$ and $f_{B_s}$.

\begin{figure}
\begin{center}
\includegraphics*[width=7.0cm,height=9.0cm,angle=270]{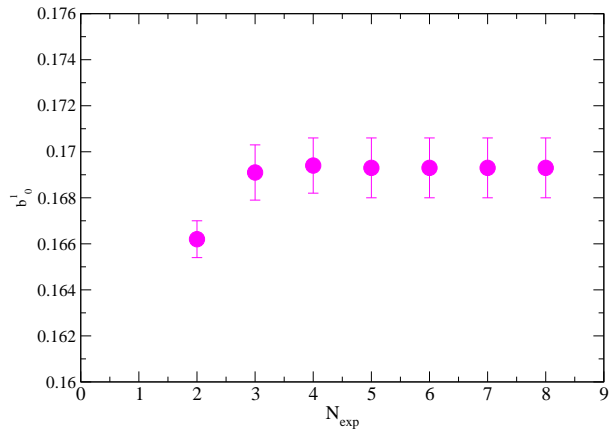}
\end{center}
\caption{
Fit results for the $B$ meson amplitude $b_0^1$ on ensemble C1 versus the 
number of exponentials $N = \tilde{N} \equiv N_{exp}$. Simultaneous $3 \times 2$ 
matrix fits were carried out to both $B$ and $B_s$ meson correlators at the same time.
 }
\end{figure}

 In all our fits we use Bayesian methods \cite{bayse} and work with fixed
 $t_{min}$ and $t_{max}$ while increasing 
the number of exponentials $N$ and $\tilde{N}$ in eq.(\ref{twopnt}) until 
fit results including errors and chisquares/dof have saturated.  Fig.6 shows fit 
results for the $B$ meson amplitde 
 $b_0^1$ on ensemble C1 versus $N$ (which we also set equal to $\tilde{N}$).
One sees that things have stabilized by $N = 4$.
In Table III we collect fit results for $ a^{3/2} \Phi \equiv a^{3/2} f_B \sqrt{M_B}$ 
and $ a^{3/2} \Phi_s \equiv a^{3/2} f_{B_s} \sqrt{M_{B_s}}$ for the six ensembles.  
The quantities $\Phi^{(0)}$ and $\Phi_s^{(0)}$ are analogous results if only the 
$ 1 \times \langle J_0^{(0)} \rangle $ contribution is included on the RHS of 
eq.(\ref{a0}), i.e. if one drops all one-loop and 1/M current corrections.
In Table IV we summarize results for the mass difference $\Delta M \equiv M_{B_s} - M_B$ 
in GeV's and the ratios $\Phi_s/\Phi$ and $\Phi_s^{(0)}/\Phi^{(0)}$.  
Fig.4 illustrates the results for $\Delta M$.  For the ratios one sees good 
agreement between $\Phi_s/\Phi$ and $\Phi^{(0)}_s/\Phi^{(0)}$ indicating complete 
lack of sensitivity to ${\cal O}(\alpha)$ or ${\cal O}(1/M)$ current corrections 
in this ratio.

\vspace{.1in}
\begin{table}[t]
\caption{ 
$\Phi \equiv f_B \sqrt{M_B}$ and $\Phi_s \equiv f_{B_s} \sqrt{M_{B_s}}$ 
in lattice units.  The lowest order results $\Phi^{(0)}$ and 
$\Phi_s^{(0)}$ are also shown. Errors include statistical and fitting 
errors.
}
\begin{center}
\begin{tabular}{|c|c|c|c|c|}
\hline
Set & $a^{3/2}\Phi^{(0)}$  &  $a^{3/2} \Phi$  &  $a^{3/2} \Phi_s^{(0)}$
  &  $a^{3/2} \Phi_s$ \\
\hline
\hline
C1  &  0.2394(18)  &  0.2214(16)   & 0.2708(13)   &  0.2508(12) \\
\hline
C2  &  0.2498(18)  &  0.2313(17)   &  0.2780(9)   &   0.2577(8)  \\
\hline
C3  & 0.2545(28)  &  0.2356(26)  &  0.2742(14)   &  0.2539(13) \\
\hline
\hline
F0  &  0.1431(16)  &  0.1293(14)  & 0.1647(8)   &  0.1489(7)   \\
\hline
F1  &  0.1483(12)  &  0.1340(11)  &  0.1664(6)   &  0.1506(5)  \\
\hline
F2  &  0.1520(12)   &  0.1375(10)  &  0.1656(7)  &  0.1498(6)  \\
\hline
\end{tabular}
\end{center}
\end{table}

\vspace{.1in}
\begin{table}
\caption{ 
  $B_s$ and $B$ mass difference $\Delta M$,
 and  ratios $ \Phi^{(0)}_s/\Phi^{(0)}$ and
$ \Phi_s/\Phi$, with $\Phi \equiv f_{B_q} \sqrt{M_{B_q}}$.
Errors include statistical and fitting errors.
}
\begin{center}
\begin{tabular}{|c|c|c|c|}
\hline
Set & $\Delta M$ [GeV]   & $\Phi_s^{(0)}/\Phi^{(0)}$ &  $\Phi_s / \Phi$\\
\hline
\hline
C1  &   0.0648(22)    &  1.1311(90)   & 1.1324(89)  \\
\hline
C2  &  0.0577(18)     &  1.1132(65)    & 1.1143(64)  \\
\hline
C3  &  0.0413(20)     &  1.0772(81)   & 1.0775(80)  \\
\hline
\hline
F0  & 0.0717(29)      & 1.1508(123)    &  1.1516(121)  \\
\hline
F1  &  0.0614(20)     &  1.1223(75)   &  1.1234(73) \\
\hline
F2  &  0.0478(13)     &  1.0889(51)   &  1.0896(50)  \\
\hline
\end{tabular}
\end{center}
\end{table}

\section{ Extracting continuum and Chiral limit physics }

In this section we describe how we extract continuum and chiral limit
 physics from  $\Phi$, $\Phi_s$ and 
$\Phi_s / \Phi$ given in Tables III and IV. 
We fit $\Phi$ and $\Phi_s$ to the general form,
\be
\label{chiral}
\Phi_q = \Phi_{0} \; (1 + \delta f_q + [analytic] ) \; ( 1 + [discret.]),
\ee
where $\delta f_q$ includes the chiral logarithm terms.  Explicit expressions, taken
 from the literature \cite{cacb,fermimilc}, 
are given in the appendix.
The chiral limit corresponds to $m_l/m_s \rightarrow (m_l/m_s)_{physical} = 1/27.4$
together with 
$m_s/m_c \rightarrow (m_s/m_c)_{physical} = 1/11.85$.
  Most of our extrapolations employed 
 formulas for $\delta f_q$ at one-loop order in chiral 
perturbation theory (ChPT) and at lowest order in 1/M.  We have also included 
some 1/M corrections such as  effects of the $B_q^*$ - $B_q$ 
hyperfine splitting as discussed in \cite{fermimilc}. 
 For the $[analytic]$ terms we use powers of 
$m_{valence}/m_c$ and $m_{sea}/\tilde{m}_c$,
 where $m_c$ is the bare HISQ charm quark mass (see
Table II) 
fixed for each ensemble through the $\eta_c$ mass, 
and $\tilde{m}_c$ is the analogous bare AsqTad charm quark mass for 
$\eta_c$ mesons made out of AsqTad quarks and antiquarks. 
The bare charm quark mass is a convenient 
scale to use since ratios such as $m_s/m_c$ or $m_l/m_c$  are equal to the 
corresponding ratio of $\overline{MS}$  masses  and are 
furthermore scale independent (up to discretization corrections). 
The ratio $\tilde{m}_c/m_c$ was found to be 0.9 in reference \cite{hisq} 
for the fine ensembles. The same ratio will be approximately true for the
 coarse ensembles as well, since $am_c$ does not vary too much for
 the lattice spacings employed here and mass 
renormalization starts only at order $\alpha_s^2$, with the one-loop 
corrections being very similar in the two actions. 

For the $[discret.]$ terms in (\ref{chiral}) we employ powers of 
$(a/r_1)^2$.  We allow for the expansion 
coefficients to be themselves functions of $aM_b$ and/or 
$am_q$ to take into account that we are dealing with an effective 
NRQCD theory for the $b$-quarks and with taste breaking splittings in 
staggered meson masses.
With NRQCD  $b$-quarks we cannot naively set 
$a \rightarrow 0$.  What we do instead is fit the data to a theoretically 
motivated ansatz for discretization errors and then remove the latter. 
For instance with our current NRQCD action the leading order discretization 
errors go as $a^2$ times a slowly varying function of $aM_b$.  Reference 
\cite{gregory} describes how we parameterize such $aM_b$ dependence.  This 
approach has worked well not just in the heavy-light spectroscopy calculations 
of \cite{gregory} but also in recent HPQCD studies of $\Upsilon$ physics 
with an improved NRQCD action \cite{upsilon}.
In the present article we have tried ans\"atze for $[discrete.]$ with both 
constant and $aM_b$-dependent coefficients mutiplying powers of $(a/r_1)$ 
and find little difference.  This corresponds to test number 6 described below. 

\begin{figure}
\begin{center}
\includegraphics*[width=8.0cm,height=9.0cm,angle=270]{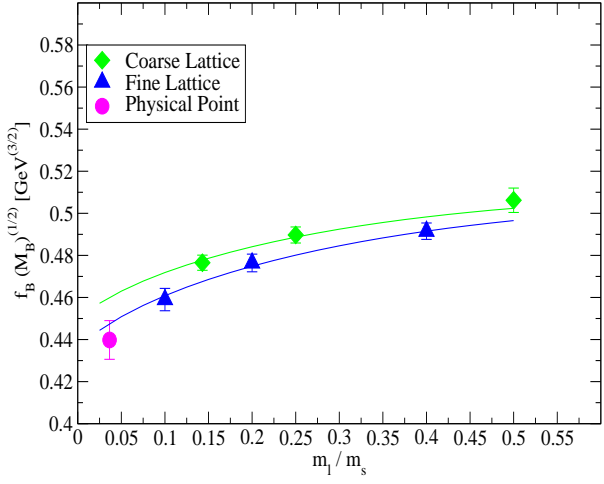}
\end{center}
\caption{
Physical point extraction for $\Phi = f_B \sqrt{M_B}$.
 }
\end{figure}

\begin{figure}
\begin{center}
\includegraphics*[width=8.0cm,height=9.0cm,angle=270]{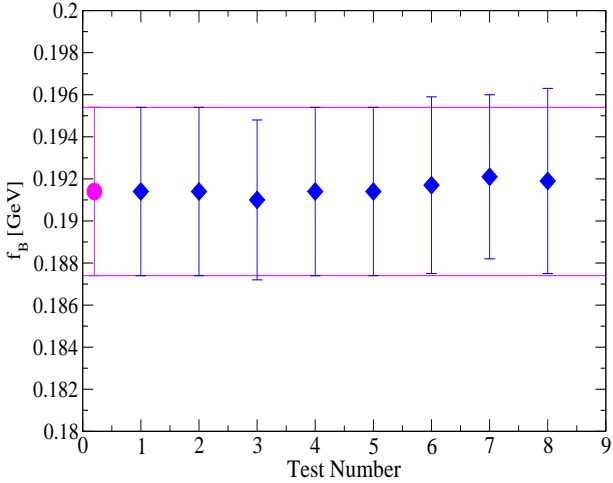}
\end{center}
\caption{
Tests of $f_B$ at the physical point.  The left most magenta point is 
the ``basic ansatz'' result.  The remaining points refer to results 
when the basic ansatz was modified in several ways as explained 
in the text.
 }
\end{figure}

\begin{figure}
\begin{center}
\includegraphics*[width=8.0cm,height=9.0cm,angle=270]{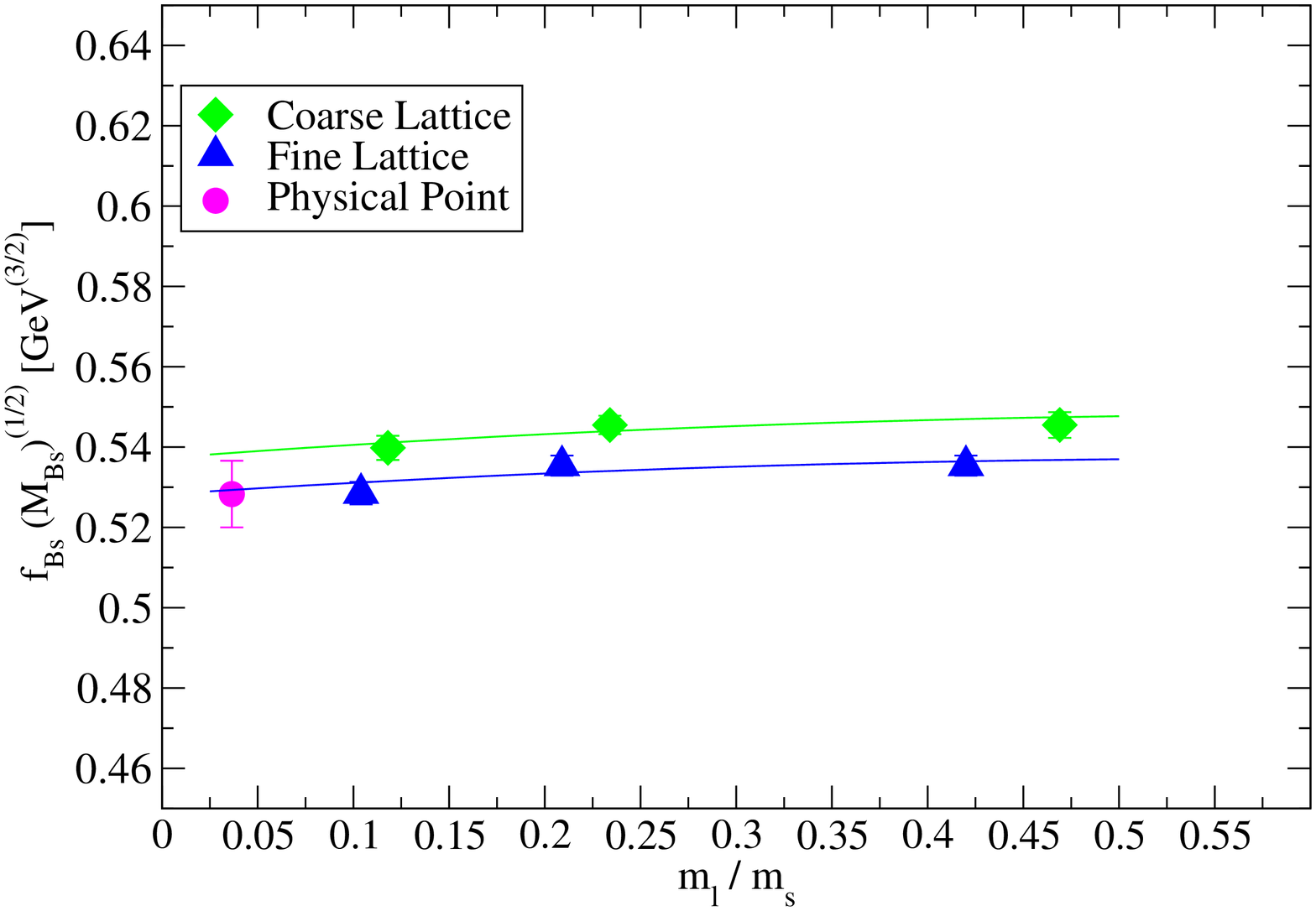}
\end{center}
\caption{
Physical point extraction for $\Phi_s = f_{B_s} \sqrt{M_{B_s}}$.
 }
\end{figure}

\begin{figure}
\begin{center}
\includegraphics*[width=8.0cm,height=9.0cm,angle=270]{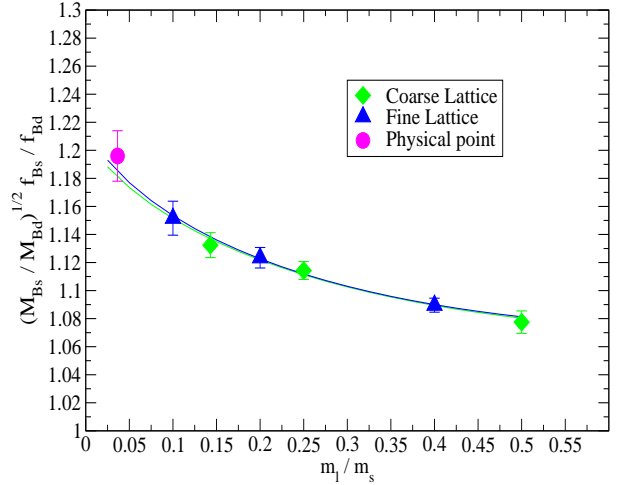}
\end{center}
\caption{
Physical point extraction for $\Phi_s/\Phi$.
 }
\end{figure}

Fig.7  shows extraction of the physical point value (the magenta point) for 
$\Phi_B$.  We show results using what we call our ``basic'' ansatz with,
\begin{eqnarray}
\label{analyt}
 &&[analytic] = \nonumber \\
&& \beta_0 (2 m_u + m_s)/\tilde{m}_c \; + \; \beta_1 m_q/m_c \; + \; 
\beta_2 (m_q/m_c)^2, \nonumber \\
\end{eqnarray}
where $m_u$($m_q$) denotes the sea(valence) light quark mass, 
\be
\label{discrete}
[discret.] = c_0 (a/r_1)^2 \; + \; c_1 (a/r_1)^4,
\ee
and using (A.7) from the appendix for $\delta f_q$. The $\chi^2/dof$ for this 
fit was 0.24.  
 We have checked the stability
of our extractions by modifying the basic ansatz in the following way:

\begin{enumerate}
\item dropping the $\beta_2$ term in (\ref{analyt});
\item  adding more $(m_q/m_c )^n$ terms with $n> 2$;
\item dropping the $c_1$ term in (\ref{discrete});
\item adding more $(a/r_1)^n$ terms with $n > 4$;
\item  making the coefficients $c_i$ depend on ´$m_q$ i.e.
a power series in $m_q/m_c$;
\item  making the coefficients $c_i$ depend on $aM_b$;
\item using (A.1) rather than (A.7) for $\delta f_q$;
\item allowing for a 20\% error in the scale $f = f_\pi$ (see appendix A 
for the relevant formulas).
\end{enumerate}
Fig.8 summarizes results from these tests. We compare $f_B$ at the 
physical point with these modifications in place with results obtained with 
the basic ansatz. The latter corresponds to the left most data point 
in Fig.8 and is the same as the magenta point in Fig.7. The integers on the 
horizontal axis in Fig.8 refer to the type of modification of the basic 
ansatz  as enumerated above. One sees that the basic ansatz result is very 
stable. The decay constant
$f_B$ changes by less than 1MeV in all the tests undertaken.

In Fig.9 and Fig.10 we show physical point extractions for 
$\Phi_s = f_{B_s} \sqrt{M_{B_s}}$ and $\Phi_s/\Phi$ both 
carried out and tested along similar lines as for $f_B$ in Fig.7 and Fig.8. 
The $\chi^2/dof$ for the two fits were 0.59 and 0.48 respectively.

The physical point results in Fig.7 and Fig.9 show statistical, extrapolation 
and $r_1^{(3/2)}$ errors whereas in Fig.10 only statistical and extrapolation errors 
are included.  In the next section we will
 discuss additional systematic uncertainties inherent in our decay 
constant determinations.
\section{Results}
\begin{table}
\caption{
Error budget }
\begin{center}
\begin{tabular}{|c|c|c|c|}
\hline
 Source  & $f_{B_s}$ &  $f_B$ &  $f_{B_s} / f_B$  \\
        &  (\%)  &  (\%) & (\%)  \\
\hline
\hline
statistical   &  0.6 &  1.2 & 1.0\\
scale $r_1^{3/2}$  &  1.1 & 1.1&  ---\\
discret. corrections &  0.9 & 0.9 &  0.9\\
chiral extrap. \& $g_{B^*B\pi}$  &  0.2 & 0.5&  0.6 \\
mass tuning    &  0.2 & 0.1 &   0.2\\
finite volume & 0.1  & 0.3  & 0.3 \\
relativistic correct.  & 1.0 & 1.0& 0.0\\
operator matching &  4.1 & 4.1&  0.1\\
\hline
 Total  &  4.4 &  4.6&  1.5\\
\hline
\end{tabular}
\end{center}
\end{table}

Table V gives the error budget for $f_B$, $f_{B_s}$ and 
$f_{B_s}/f_B$.  The first four entries, ``statistical'', ``scale $r_1^{3/2}$'', 
 ``discret. corrections'' and ``chiral extrapolation \& $g_{B^*B\pi}$'' are 
all part of the errors emerging automatically from the fits. 
 Their individual contributions  were separated out 
 using the methods of reference \cite{alphas} (see eq.(30) and (31) of that article).
The remaining four entries in Table V, ``mass tuning'', 
``finite volume'', ``relativistic corrections'' and ``operator matching'' are 
additional systematic errors affecting our calculations.  Sensitivity to the strange 
quark mass can be estimated by comparing results for valence quarks masses 
$am_s$ and $am_l$.  Similarly effects of mistuning of $aM_b$ can be investigated 
using older NRQCD/HISQ decay constant results (see \cite{lat09})
 covering a range of $aM_b$ values.  Those calculations were done 
before proper retuning of the $b$-quark mass and provide information on how
 the decay constants depend on $aM_b$.
For the ``finite volume'' uncertainty we take the same percentages as determined 
for the $D$ and $D_s$ meson decay constants in \cite{fdprl} using finite volume chiral 
perturbation theory.  Our heavy-light currents have been matched to 
full QCD through order $  \alpha_s \Lambda_{QCD}/M_b $ and corrections come in 
at order $ ( \Lambda_{QCD}/M_b )^2 \approx 0.01 $.  There are order 
$ \alpha_s \Lambda_{QCD}/M_b $ corrections to the NRQCD action that are not 
included in our simulations. However, as discussed in \cite{bmixing}, their effect on 
decay constants can be bounded to be at most $\sim 1$\%. 

The $O(\alpha_s^2)$ corrections to eqs.(\ref{a0}-\ref{jsub})
 are not known. The $J_0^{(i),sub}$ are
 nonleading, so the most important high-order correction is in the coefficient of $J_0^{(0)}$.
 To account for corrections at this level and beyond, we modify our data
 by multiplying the right-hand side of eq.(\ref{a0}) by an overall factor of 
$1 + \alpha^2_s \rho_0^\prime$ where we approximate $\alpha_s^2\approx0.1$. 
We use two different $\rho_0^\prime$s, one for all coarse-lattice data and the other
 for all fine-lattice data. To be conservative, we take each to be $O(0.4)$, which
 is more than twice as large as the one-loop 
$\rho_0$ and also comparable to the estimates used in \cite{bmixing}: 
that is, we set each $\rho_0^\prime = 0\pm0.4$.
 The errors from these factors are combined in quadrature with the simulation
 errors in the currents, taking care to preserve the correlations caused by the fact
 that all course-lattice data has the same $\rho_0^\prime$, as does all fine-lattice data.
We then repeat the fits to eq.(\ref{chiral}) described in the previous 
section, this time applied to the modified data with enhanced errors.
 We use the difference 
between the total extrapolation
 error obtained with and without higher order matching errors added to the 
data to estimate the operator matching errors for $f_B$ and $f_{B_s}$. These are given 
as the last entries in Table V.
 For the ratio $f_{B_s}/f_B$, 
matching errors are negligible, as was already pointed out at the end of 
section IV.

 Finally, we note that 
sea charm quarks are omitted in our simulations. However we expect their 
contributions to be small enough  that the final total errors in 
Table V are unaffected.

Our final decay constant results including all the errors discussed above 
are,
\be
\label{fbres}
f_B = 0.191(9) {\rm GeV},
\ee
\be
\label{fbsres}
f_{B_s} = 0.228(10) {\rm GeV},
\ee
and
\be
\label{ratres}
\frac{f_{B_s}}{f_B} = 1.188(18).
\ee
These numbers are in good agreement with HPQCD's previous NRQCD $b$-quark/AsqTad 
light quark results \cite{bmixing},
 however with improved total errors. Comparison plots 
are shown in the next section.

\begin{figure}
\begin{center}
\includegraphics*[width=8.0cm,height=12.0cm,angle=270]{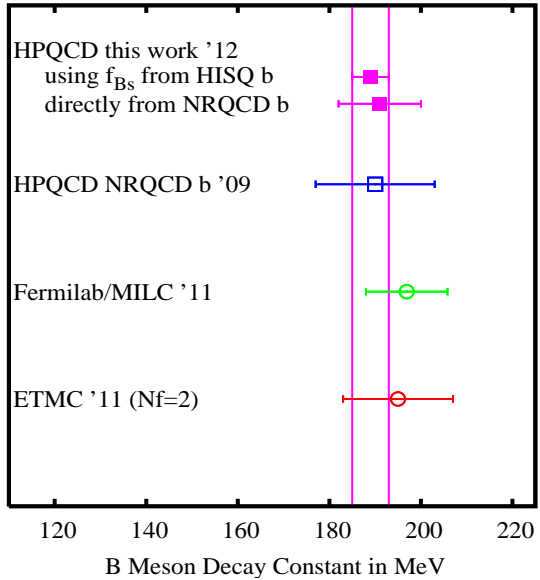}
\end{center}
\caption{
Comparisons of results for $f_B$ 
from this article with previous HPQCD work \cite{bmixing} and with 
results from  
the Fermilab/MILC \cite{fermimilc} and ETM \cite{etmc} collaborations.
 }
\end{figure}

\begin{figure}
\begin{center}
\includegraphics*[width=8.0cm,height=12.0cm,angle=270]{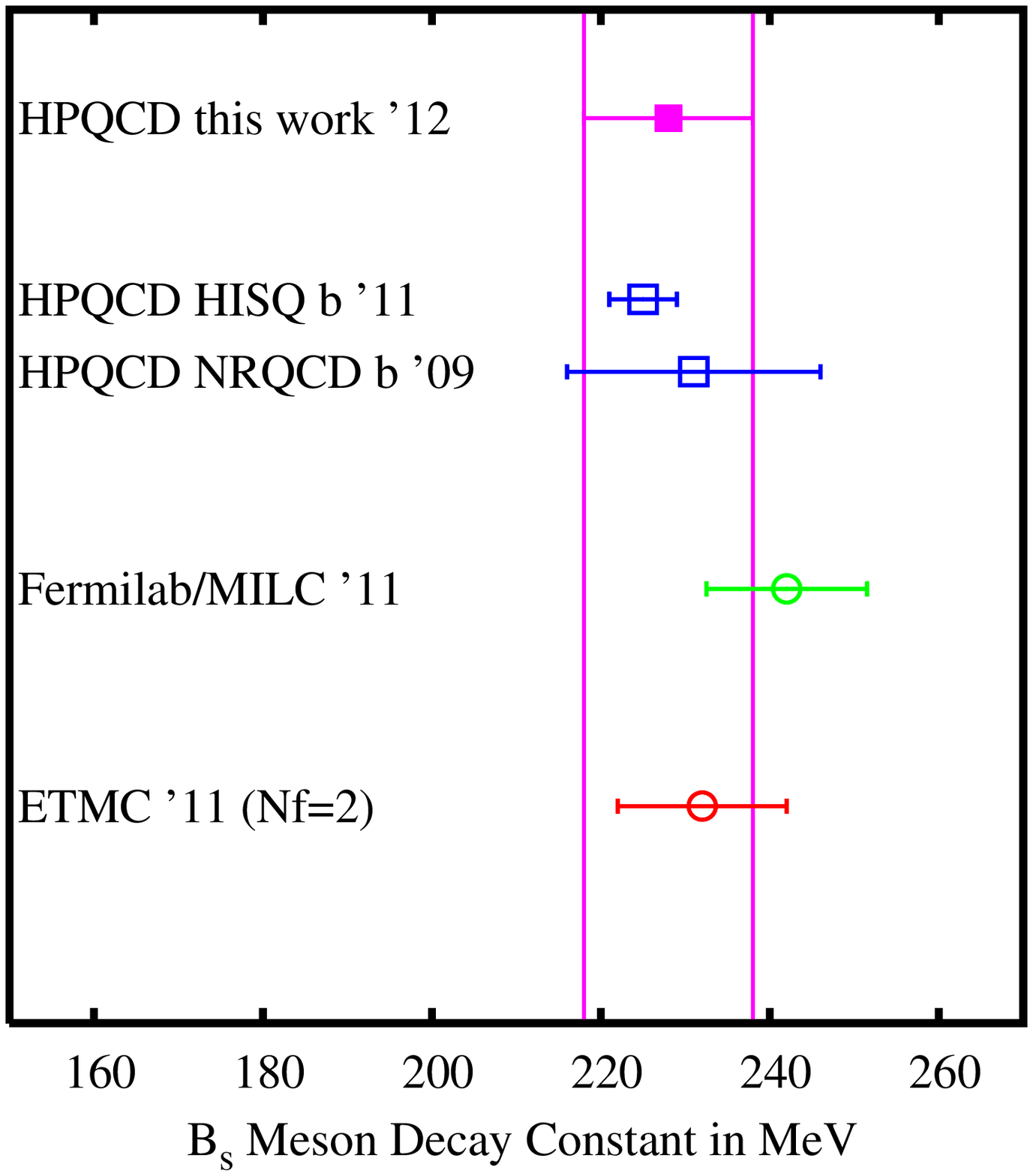}
\end{center}
\caption{
Comparisons of results for $f_{B_s}$ 
from this article with previous HPQCD work \cite{bmixing,hisqfbs} and with results from 
the Fermilab/MILC \cite{fermimilc} and ETM \cite{etmc} collaborations.
 }
\end{figure}

\begin{figure}
\begin{center}
\includegraphics*[width=8.0cm,height=12.0cm,angle=270]{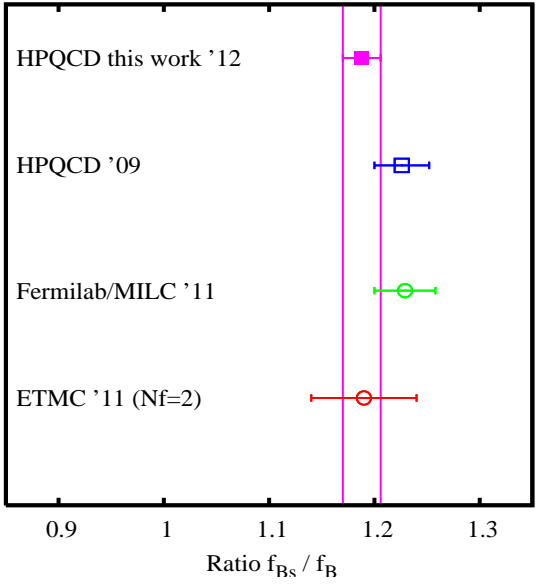}
\end{center}
\caption{
Comparisons of results for $f_{B_s}/f_B$ 
from this article with previous HPQCD work \cite{bmixing} and with results from 
the Fermilab/MILC \cite{fermimilc} and ETM \cite{etmc} collaborations.
 }
\end{figure}

The errors in $f_B$ and $f_{B_s}$ are 
overwhelmingly dominated by the matching uncertainties.  Without them, the 
total errors would be reduced to $4.6$\% $\rightarrow 2.1$\% and 
$4.4$\% $\rightarrow 1.6$\% for $f_B$ and $f_{B_s}$ respectively. Clearly a huge 
advantage can be gained if one could develop a formalism that did not 
require operator matching.  One major motivation for designing the 
HISQ action \cite{hisq} was to come up with a quark action that could be used not 
only for accurate light quark physics, but also to simulate heavy quarks.  
It has been employed already very succesfully for charmed quarks
 \cite{fdprl, dtok, dtopi, fds2} and 
HPQCD has recently also started work with $am_Q > am_c$ \cite{hisqfbs}.
  The HISQ action 
allows for a relativistic treatment of heavy quarks which means that one does not  
have to resort to effective theories.  One important consequence is that 
decay constants can be determined from 
absolutely normalized currents.  There is no need for operator matching. 
Furthermore it has been demonstrated that due to its high level of 
improvement the HISQ action can be used for heavy quarks up to about
$am_Q \leq 0.8$ without leading to large discretization effects. Recently 
a succesful application of heavy HISQ quarks to $B$ physics was achieved 
through a very accurate determination of the $B_s$ meson decay 
constant, namely $f^{(HISQ)}_{B_s} = 0.225(4)$GeV
 with errors of only $1.8$\% \cite{hisqfbs}.
There is very good agreement between $f^{(HISQ)}_{B_s}$ and the 
NRQCD $b$-quark result eq.(\ref{fbsres}) of this article.  This indicates 
that the very different systematic errors in the two calculations are under 
 control and properly accounted for in our error estimates.

\begin{figure}
\begin{center}
\includegraphics*[width=8.0cm,height=12.0cm,angle=270]{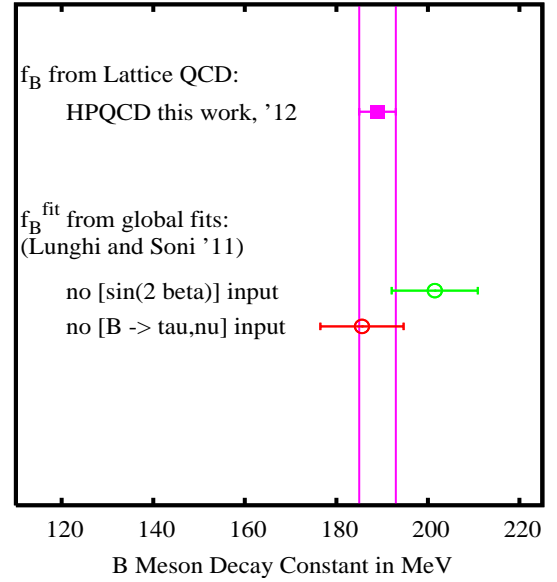}
\end{center}
\caption{
Comparisons with $f_B^{(fit)}$ from global fits as given in \cite{ls2}.
 }
\end{figure}

 The HISQ $b$-quark calculation of $f_{B_s}$ required going to very fine 
lattices including the MILC superfine and ultrafine ensembles with 
lattice spacings $\sim 0.06$fm and $\sim 0.045$fm respectively.  Repeating those 
calculations for the $B$ meson with its light valence quark would be quite 
expensive and it will take some time before such calculations become 
available. In the mean time we can combine $f^{(HISQ)}_{B_s}$ with 
the result eq.(\ref{ratres}) of this article to extract a new and 
accurate $f_B$.  One finds,
\be
\label{fbres2}
\left [ \frac{f_{B_s}}{f_B} \right ]^{-1}_{NRQCD} \times f^{(HISQ)}_{B_s}
 \equiv f_B = 0.189(4) {\rm GeV},
\ee
which is in excellent agreement with (\ref{fbres}), only more accurate by better 
than a factor of 2.
Eq.(\ref{fbres2}) is the most important result of this article for phenomenology. 
It also demonstrates the advantages of working with both HISQ and NRQCD 
$b$-quarks in parallel. In the future we plan to apply this combined approach to 
$B$ and $B_s$ semileptonic decay studies as well.

\section{Summary}

We have carried out new determinations of $f_B$, $f_{B_s}$ and 
$f_{B_s}/f_B$ using NRQCD $b$-quarks and HISQ light valence quarks and improve 
on our previous calculations with AsqTad light quarks.  Figures 11, 12 and 13 
compares our new results with HPQCD's older work \cite{bmixing, hisqfbs} and also 
with results from the Fermilab/MILC \cite{fermimilc}
 and the ETM \cite{etmc} collaborations. 
One finds overall consistency between the different lattice groups. 
Our most accurate determination of $f_B$, eq.(\ref{fbres2}), comes from 
combining the new ratio $f_{B_s}/f_B$, (\ref{ratres}), with 
a precise determination of $f_{B_s}$ based on HISQ $b$-quarks \cite{hisqfbs}.
 This gives the 
most precise $f_B$ available today with errors of just $2.1$\%.  As mentioned 
in the introduction, accurate values for  $f_B$ are
needed to compare with $f_B^{(fit)}$ 
from global fits in Unitarity Triangle analyses. In Fig.14 we compare the new 
 accurate $f_B$ with two examples of $f_B^{(fit)}$ determined by 
Lunghi and Soni \cite{ls2}. With current errors the two $f_B^{(fit)}$ values 
are consistent with each other and with $f_B$ from Lattice QCD. In the 
future, once errors are reduced considerably, these kind of comparisons 
could become more interesting.

{\bf Acknowledgements}: \\
This work was supported by the DOE (DE-FG02-91ER40690, 
DE-FG02-04ER41302, DE-AC02-06CH11357), 
the NSF (PHY-0757868), and the STFC. Numerical simulations were
 carried out on facilities
of the USQCD collaboration funded by the Office of Science of the DOE and 
at the Ohio Supercomputer Center.  Some calculations for this work were 
performed on the DiRAC facility jointly funded by the STFC and BIS.  
We thank the MILC collaboration for use 
of their gauge configurations.

\appendix
\section{Partially Quenched ChPT Chiral Logs}
In this appendix we summarize partially quenched ChPT (PQChPT) expressions for the 
chiral logarithm terms $\delta f_B$ and $\delta f_{B_s}$,
taken from the literature.
 We follow closely the notation of \cite{cacb} which we also adopted in our 
$D \rightarrow K$ semileptonic paper \cite{dtok}.
 We use ``$u$'' and ``$s$'' for sea and ``$q$'' and ``$q_s$'' for valence 
light and strange quarks respectively.
Furthermore $m_{ab}$ is the mass of the pseudoscalar meson 
with quark/antiquark content $a$ and 
$b$ and $m^2_\eta = \frac{1}{3}(m^2_{uu} + 2 m^2_{ss})$. 
For $x = q$ or $q_s$ PQChPT gives,
\begin{eqnarray}
\label{pqchpt}
\delta f_{B_x} &=& 
\frac{1 + 3 g^2}{32 \pi^2 f^2} \left [ -2 I_1(m_{xu}) - I_1(m_{xs}) \right. 
\nonumber \\
&&  \left. - 
\frac{1}{3} DR^{[2,2]}(m_{xx},I_1) \right ] ,
\end{eqnarray}
where
\be 
I_1(m) = m^2 log\frac{m^2}{\Lambda^2},
\ee
and
\be
DR^{[2,2]}(m;{\cal I}) = \frac{\partial}{\partial m^2} R^{[2,2]}(m; {\cal I}),
\ee
with
\begin{eqnarray}
R^{[2,2]}(m;{\cal I}) &=& \frac{(m^2_{uu} - m^2)(m^2_{ss} - m^2)}{(m^2_\eta - m^2)} 
{\cal I}(m) \nonumber \\
& +& \frac{(m^2_{uu} - m^2_\eta)(m^2_{ss} - m^2_\eta)}{(m^2 - m^2_\eta)} 
{\cal I}(m_\eta).  \nonumber \\
\end{eqnarray}
In eq.(A1) $g$ is the 
$B^*B\pi$ coupling which has not been measured yet experimentally, 
but for which several unquenched lattice determinations are now available 
\cite{gsq}. 
We treat this ``constant'' as one of the fit 
parameters and set  priors for the square of this coupling to a
  central value of $g^2 =  0.25$ with 
width 0.10 (40\%). This is consistent
 with typical values in the recent literature \cite{gsq}. 
The scale $\Lambda$ is set to $4 \pi f$, with $f$  given by 
 the physical pion decay constant.
In the full QCD limit the partially quenched formulas simplify to,
\be
\delta f_{B_s} = 
\frac{1 + 3 g^2}{32\pi^2 f^2} \left [ -2 I_1(m_K) -  \frac{2}{3}I_1(m_\eta) \right],
\ee
\begin{eqnarray}
\delta f_B &=& 
\frac{1 + 3 g^2}{32 \pi^2 f^2} \left [ -\frac{3}{2} I_1(m_\pi) -  \frac{1}{6}I_1(m_\eta) 
 - I_1(m_K) \right].\nonumber \\
\end{eqnarray}
Following \cite{fermimilc} we have also considered PQChPT logs that 
include hyperfine and flavor splittng effects.  A modification of the 
terms proportional to $3g^2$ in (\ref{pqchpt}) is required leading to,
\begin{eqnarray}
\label{pqchpt2}
\delta f_{B_x} &=& 
\frac{1 }{32 \pi^2 f^2} \left [ -2 I_1(m_{xu}) - I_1(m_{xs}) \right. 
\nonumber \\
&&  \left. - 
\frac{1}{3} DR^{[2,2]}(m_{xx},I_1) \right ] + \nonumber \\
 & & \frac{3 g^2 }{32 \pi^2 f^2} \left [ -2 J(m_{xu},\Delta + \delta_{xu}) 
- J(m_{xs},\Delta + \delta_{xs}) \right.   \nonumber \\
&&  \left. - 
\frac{1}{3} DR^{[2,2]}(m_{xx},J(m_{xx},\Delta)) \right ], \nonumber \\
\end{eqnarray}
with,
\be
J(m,\Delta) = (m^2 - 2 \Delta^2) \; log \left ( \frac{m^2}{\Lambda^2} \right ) 
+ 2 \Delta^2 - 4 \Delta ^2 F(m/\Delta),
\ee
\be
F(1/x) = \cases{ - \frac{ \sqrt{1 - x^2}}{x} \; \left [
\frac{\pi}{2} - tan^{-1} \frac{x}{\sqrt{1 - x^2}} \right ] \qquad
 |x| \leq 1 \cr
                    \cr
              \frac{\sqrt{x^2 -1 }}{x} \; log(x +  \sqrt{x^2 -1})
\qquad  \qquad |x| \geq 1 \cr }
\ee
$\Delta$ is the $B_x^* - B_x$ hyperfine splitting and $\delta_{xu}$ and 
$\delta_{xs}$ adjust for the fact that in some one-loop diagrams the 
internal $B^*_{u/s}$ does not have the same flavor as the external 
$B_x$.
We have carried out chiral/continuum extrapolations with both 
(A.1) and (A.7).  Differences in the final values at the physical 
point serve as a measure of systematic errors coming from our extrapolation 
ansatz. 

\section{Example of priors used in section V}
Table VI gives a sample set of priors and prior widths used for the 
$f_B$ extraction in section V. 
For parameters such as $\beta_j$ or $c_j$ where the overall sign is not 
known a priori, we take the central value to be 0.0.
 The widths for the $\beta_j$ depend on whether 
$m_c$, $1/a$ or $1/r_1$ is used to set the scale for the masses. Although $m_c$ 
is our prefered scale, due to the ease of handling quark mass running issues, 
we have also tried fits with the other scales and obtain consistent results. 
In all cases fitted values for the parameters are consistent and within the 
widths assigned to them.  For $c_0$ we use prior widths of 0.3 to reflect 
the expectation that ${\cal O}(a^2)$ errors come in as ${\cal O}(\alpha_s a^2)$.
Again fit results for $c_0$ are consistent with this expectation.

\begin{table}
\caption{
Priors and prior widths for fits to eq.(\ref{chiral}) }
\begin{center}
\begin{tabular}{|c|c|c|}
\hline
  &  prior & width  \\
\hline
\hline
$\Phi_0$  &  1.00    &  1.00  \\
$\beta_0$ &  0.00    &  1.00  \\
$\beta_1$  &  0.00    &  4.00  \\
$\beta_2$  &  0.00  &  1.00  \\
$c_0$     &  0.00  &   0.30  \\
$c_1$     &  0.00  &   1.00 \\
$g^2$  &   0.25   &   0.10 \\
$r_1$  &  0.3133  &   0.0023 \\
\hline
\end{tabular}
\end{center}
\end{table}


\begin{thebibliography}{99}
\bibitem{mixed}
This article is part of HPQCD's program to study $D$, $D_s$, $B$ and
      $B_s$ decays and mixing on the MILC $N_f=2+1$
      lattices with HISQ valence light, strange and charm quarks.
      Results for $f_D$, $f_{D_s}$, $B$, $B_s$ and $B_c$ spectroscopy, 
$m_s$, $m_c$, $m_b$, $\alpha_s$, and
    for  $D$ semileptonic form factors and $V_{cd}$ and $V_{cs}$ have 
already been published.  This mixed action (HISQ valence on AsqTad sea) 
 approach has also been tested 
through precision studies of light quark physics quantities such 
as $f_\pi$ and $f_K$, where excellent agreement with experiment 
and results from non-mixed-action calculations (such as AsqTad on Asqtad) 
was found.
\end{thebibliography}


\end{document}